\documentclass[journal]{IEEEtran}
\usepackage{amsmath}
\usepackage{graphicx}
\begin{document}

\title{On the Maximum Sum-rate Capacity of Cognitive Multiple Access Channel}

\author{Peng Cheng, Guanding Yu$^\dag$, Zhaoyang Zhang, Hsiao-Hwa Chen, and Peiliang Qiu
\thanks{This work was supported by the National Natural Science Foundation of China (No.
60472079, 60572115).}\thanks{P. Cheng, G. Yu, Z. Zhang and P. Qiu
are with the Institute of Information and Communication
Engineering, Zhejiang University, Hangzhou, China.}\thanks{H. Chen
is with the National Sun Yat-Sen University,
Taiwan.}\thanks{$^\dag$Corresponding author. Email:
yuguanding@zju.edu.cn}} \maketitle

\begin{abstract}
We consider the communication scenario where multiple cognitive
users wish to communicate to the same receiver, in the presence of
primary transmission. The cognitive transmitters are assumed to
have the side information about the primary transmission. The
capacity region of cognitive users is formulated under the
constraint that the capacity of primary transmission is not
changed as if no cognitive users exist. Moreover, the maximum
sum-rate point of the capacity region is characterized, by
optimally allocating the power of each cognitive user to transmit
its own information.
\end{abstract}


\section{Introduction}
With the increasing demand of radio spectrum, Cognitive Radio (CR)
\cite{overview1} has drawn great attention in the world these days
for its ability of operation in licensed bands without a license.
In cognitive radio systems, the cognitive (unlicensed) user needs
to detect the presence of the primary (licensed) users as quickly
as possible and dynamically change system parameters, such as the
transmitting spectrum and modulation schemes, so as to best
utilize the valuable spectrum.

Recently, several papers \cite{info1},\cite{info2} have discussed
the achievable rate of such a cognitive radio system from the view
of information theory. Literature \cite{info2} studied the
achievable rate of cognitive user under the constraint that (i) no
interference is created for the primary user, and (ii) the primary
encoder-decoder pair is oblivious to the presence of cognitive
radio. The authors pointed that, the maximum achievable rate is
achieved by a mixed strategy of dirty paper coding and cooperation
with the primary user.

In this letter, we further study the scenario containing multiple
cognitive users in the presence of primary transmission. These
cognitive users wish to communicate with a same access point (AP).
We first model this system as a cognitive multiple access channel
(MAC) and character the capacity region of the cognitive MAC
channel. Then, the problem of maximizing the sum-rate capacity is
formulated as a nonlinear optimization problem. Employing the
classic Lagrangian multiplier method, an iterative algorithm to
achieve the maximum sum-rate capacity is proposed.

The remainder of this letter is organized as follows. Section II
gives the system model. Section III abstracts the system model in
Section II into an information theory model, and the capacity
region of it is analyzed. In Section IV, the problem of achieving
the sum-rate optimal point is discussed and the corresponding
algorithm is given. Finally, we conclude the letter in Section V.
\section{System Model}
The system model we study is the heterogeneous network depicted in
Fig. 1. There is a primary user $U_p$ communicating with the Base
Station (BS) in a large 3G cellular network, which uses a licensed
spectrum. A small WLAN network without license of the spectrum
lies in the communication range of a 3G cell, where $K$ cognitive
radio users $U_i (i=1,2,..K)$ want to access the WLAN Access Point
(AP).
\begin{figure} \centering
\includegraphics[width=0.28\textwidth] {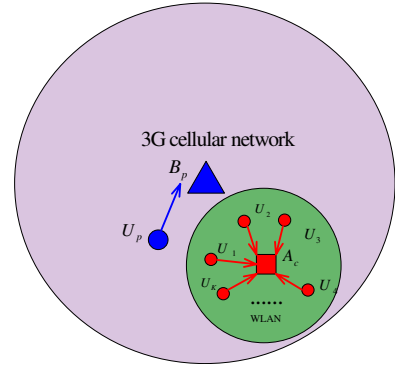}
\caption {The system model of the multiple cognitive users access}
\label{fig3}
\end{figure}

For the above scenario to be studied, we assume the primary user
$U_p$ has message $m_p\in\{0,1,...,2^{nR_p}\}$ intended for the
primary receiver $B_p$ to decode, and the cognitive user $U_i
(i=1,2,...,K)$ has message $m_i\in\{0,1,...,2^{nR_i}\}$ intended
for the cognitive receiver $A_c$. The average transmit powers of
$U_p$ and $U_i$ are constrained by $P_p$ and $P_i (i=1,2,...,K)$,
respectively. It is assumed that all cognitive users have perfect
knowledge of the primary user's codeword $m_p$.
\section{Capacity Region}
In this section, we will model the heterogeneous network in Fig. 1
as an information theory model, and then analyze its capacity
region. The information theory model of the heterogeneous network
is depicted in Fig.2. The small WLAN can be viewed as a
multiple-access channel with inference from the primary user,
whose receive signal is expressed as follows:
\begin{equation}
Y^n=\sum^K_{k=1}h_k\cdot X^n_k+f\cdot X_p+ Z^n,  \label{equ1}
\end{equation}
where $h_k,(k=1,2,...,K)$ is the path gain of the link from
cognitive user $U_i$ to $A_c$, $f$ is the path gain of the
inference link from $U_p$ to $A_c$, and $Z^n$ is additive white
Gaussian noise with variance ${\sigma_c}^2$. For the large 3G
cellular network, the received signal model is:
\begin{equation}
Y^n_p=h_p\cdot X^n_p+ \sum^K_{k=1} g_k\cdot X^n_k +Z^n_p,
\label{equ2}
\end{equation}
where $h_p$ is the path gain of the link from primary user $U_p$
to $B_p$, $g_k,(k=1,2,...,K)$ is the path gain of the inference
link from $U_i$ to $B_p$, and $Z^n_p$ is additive white Gaussian
noise with variance ${\sigma_p}^2$.

\begin{figure} \centering
\includegraphics[width=0.33\textwidth] {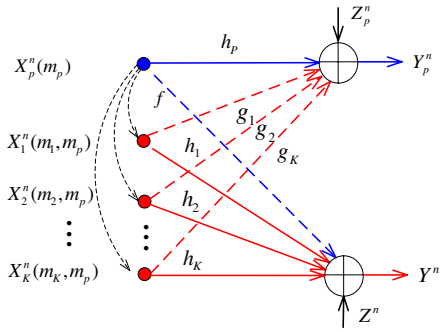}
\caption {The information theory model of the cognitive
multiple-access channel} \label{fig3}
\end{figure}
The capacity region of the cognitive multiple access channel is
defined as the set of achievable rates of cognitive users under
the constraints that (i) no interference is created for the
primary user, and (ii) the primary encoder-decoder pair is
oblivious to the presence of cognitive radio. A simpler scenario
which contains only one cognitive user is studied in \cite{info2}.
The largest rate of the cognitive user is achieved by a mixed
strategy of dirty paper coding and cooperation with the primary
user. The optimal coding strategy for the cognitive user $U_k$ is:
\begin{equation}
X_k^n  = \hat X_k^n  + \gamma _k \sqrt {\frac{{P_k }}{{P_p }}}
X_p^n , k = 1,2,...K,  \label{equ3}
\end{equation}
where the first term $\hat X_k^n$ denotes that the user $U_k$'s
codeword $m_k$ is dirty-paper coded based on the primary user's
codeword $m_p$ (so it is called the \emph{dirty-paper code part}),
and the second term, the \emph{cooperation part}, is the
duplicated information of the primary user, which is combined with
the information from $U_p$ in order to assure the primary user's
rate. In addition, $\gamma_k^2$ denotes the power ratio of the
cooperation part.

\begin{figure} \centering
\includegraphics[width=0.3\textwidth] {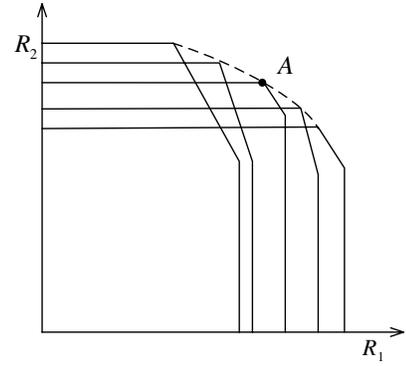}
\caption {A two-user capacity region as a union of capacity
regions, each corresponding to a feasible ${\boldsymbol \gamma}\in
\mathcal{F}$. The point A corresponds to the maximum sum-rate
point, which will be exploited in the next section.} \label{fig6}
\end{figure}

For the receiver of the AP, it can decode each cognitive user's
signal $X_k^n (k=1,2,...,K)$ as if there is no interference from
the primary user due to the dirty paper coding. Therefore, we can
regard this model as a Gaussian multiple-access channel, and its
capacity region is written as
\begin{equation}
\begin{split}
C_{mac} ({\bf h},{\bf P},{\boldsymbol \gamma }) &= \biggl\{ {\bf
R}:{\bf R}(S) \le \frac{1}{2}\log \biggl( {1 +
\frac{{\sum\limits_{k \in S} {h_k^2 (1 - \gamma _k^2 )P_k } }}{
{{\sigma_c} ^2 }}} \biggl), \\
& \phantom{{=}\biggl\{} \forall S \subset \{ 1,2,....,K\}\biggl\}
\end{split}
\label{equ5}
\end{equation}
where ${\bf h}=(h_1,h_2,...,h_K)^T$ and ${\bf
P}=(P_1,P_2,...,P_K)^T$ are constant, but $\boldsymbol{\gamma} =
(\gamma_1,\gamma_2,...,\gamma_K)^T$ is a variable vector. For the
receiver of $B_p$, its rate can be expressed as follows:
\begin{equation}
R_P  = \frac{1}{2}\log \biggl( {1 + \frac{{\biggl( {h_p\sqrt {P_p
} + \sum\limits_{k = 1}^K {g_k \gamma _k \sqrt { P_k } } }
\biggl)^2 }}{{{\sigma_p}^2 + \sum\limits_{k = 1}^K {g_k ^2 (1 -
\gamma _k^2 )P_k } }}} \biggl),  \label{equ6}
\end{equation}
which should be equal to the original rate as if there is no
interference from the cognitive users, i.e., the following
equation must be satisfied:
\begin{equation}
\frac{1}{2}\log \biggl( {1 + \frac{{h_p^2 P_p }}{{\sigma _p^2 }}}
\biggl) = \frac{1}{2}\log \biggl( {1 + \frac{{\biggl( {h_p \sqrt
{P_p } + \sum\limits_{k = 1}^K {g_k \gamma _k\sqrt { P_k } } }
\biggl)^2 }}{{\sigma _p^2  + \sum\limits_{k = 1}^K {g_k ^2 (1 -
\gamma _k^2 )P_k } }}} \biggl). \label{equ7}
\end{equation}
Define the set of feasible power allocation policies $\mathcal{F}$
as
\begin{equation}
\mathcal{F}=\biggl\{\boldsymbol{\gamma}: \frac{h_p^2{P_p
}}{{{\sigma_p} ^2 }} = \frac{{\biggl( {h_p\sqrt {P_p }  +
\sum\limits_{k = 1}^K {g_k \gamma _k\sqrt { P_k } } } \biggl)^2
}}{{{\sigma_p} ^2  + \sum\limits_{k = 1}^K {g_k ^2 (1 - \gamma
_k^2 )P_k } }}\biggl\},  \label{equ8}
\end{equation}
then the capacity region of the cognitive MAC channel is the
convex hull of all capacity regions for every ${\boldsymbol
\gamma}\in \mathcal{F}$. That is,
\begin{equation}
{\boldsymbol C} = Co\biggl \{ \mathop  \cup \limits_{{\boldsymbol
\gamma } \in \mathcal{F}} C_{mac} ({\bf h},{\bf P},{\boldsymbol
\gamma })\biggl \} \label{equ9}
\end{equation}
An example of capacity region of a two-user cognitive MAC channel
is shown in Fig. \ref{fig6}.
\section{Maximum Sum-rate Point}
In this section, our interests turn to the maximum sum rate point
on the capacity region \textbf{\emph{C}}. For a fixed
${\boldsymbol \gamma }\in \mathcal{F}$, the sum-rate capacity of
the multiple cognitive radio access network is expressed as
follows:
\begin{equation}
R_{opt} (\boldsymbol \gamma) = \frac{1}{2}\log \biggl( {1 +
\frac{{\sum\limits_{k = 1}^K {(1 - \gamma _k^2 )h_k^2 P_k }
}}{{\sigma _c^2 }}} \biggl).  \label{equ10}
\end{equation}
Now, the problem is how to optimally choose the $\gamma_k
(k=1,2,...,K)$ so as to maximize the the sum-rate capacity, which
can be formulated as a nonlinear optimization problem:
\begin{equation}
\!\!\!\!\!\! \!\!\!\!\!\!\!\!\! \!\!\! \!\!\!\!\!\! \!\!\!
\!\!\!\!\!\!  \textbf{maximize} ~~~~~~ \sum\limits_{k = 1}^K {(1 -
\gamma _k^2 )h_k ^2 P_k }, \label{equ11}
\end{equation}
~~~~~~~ \textbf{subject to}
\begin{equation} \left\{
\begin{aligned}
\frac{h_p^2P_p}{{\sigma_p}^2}  &= \frac{{\biggl({h_p\sqrt {P_p }  + \sum\limits_{k = 1}^K {g_k \gamma _k\sqrt{ P_k } } } \biggl)^2 }}{{{\sigma_p}^2 + \sum\limits_{k = 1}^K {g_k ^2 (1 - \gamma _k^2 )P_k } }} \\
0 &\le \gamma _k  \le 1,k = 1,2,...,K
\end{aligned}
\right. \label{equ12}
\end{equation}

We can see that, the optimization object is quadratic and concave,
and meanwhile, the constraint is quadratic and convex, but not
affine. So we can't solve the above optimization by classical
convex optimization algorithms \cite{cvx}.

Therefore, we will give an iterative algorithm to solve the above
problem. First, a Lagrangian multiplier is defined as
\begin{align}
J({\boldsymbol \gamma } ) &= \sum\limits_{k = 1}^K {(1 - \gamma
_k^2 )h_k ^2 P_k } + \lambda \biggl[ {{\sigma_p} ^2 \biggl(
{h_p\sqrt {P_p } + \sum\limits_{k = 1}^K {g_k \gamma_k \sqrt {P_k
} } } \biggl)^2 }\biggl. \notag \\
& -\biggl.{ h_p^2P_p \biggl( {{\sigma_p} ^2 + \sum\limits_{k =
1}^K {g_k ^2 (1 - \gamma _k^2 )P_k } } \biggl)} \biggl].
\label{equ13}
\end{align}
If there is no constraint of $0 \le \gamma _k  \le 1$, a unique
maximizer of $J({\boldsymbol \gamma })$ exists when
\begin{equation}
\frac{{dJ}}{{d\gamma _k }} =  - 2h_k ^2 P_k\gamma _k  + 2\lambda
{\sigma_p}^2Xg_k \sqrt {P_k } + 2\lambda h_p^2 P_p g_k ^2 P_k
\gamma _k = 0, \label{equ14}
\end{equation}
where
\begin{equation}
X = h_p\sqrt {P_p }  + \sum\limits_{k = 1}^K {g_k \gamma _k\sqrt {
P_k } }.  \label{equ15}
\end{equation}
Taking into account the restrict of $0\le\gamma_k\le 1$, we obtain
\begin{equation}
\gamma _k  =
 \begin{cases} \frac{\displaystyle \lambda
{{\sigma_p}^2X}}{\displaystyle {(\beta_k ^2 - \lambda h_p^2 P_p
)g_k \sqrt {P_k
} }}& \text{if $0\leq \gamma_k<1$}, \\
1& \text{else},
\end{cases}
\label{equ16}
\end{equation}
where $\beta_k  =h_k /g_k$. Define the set $\mathcal{S}$ as:
\begin{equation}
\mathcal{S}=\{k:\gamma_k<1,\forall k\}, \label{equ17}
\end{equation}
and substitute (\ref{equ16}) into (\ref{equ15}), $X$ can be
determined as
\begin{equation}
X = \frac{h_p \sqrt {P_p }  + \sum\limits_{k
\in\overline{\mathcal{S}} } g_k\sqrt {P_k }}{1 - \lambda \sigma
_p^2 \sum\limits_{k \in \mathcal{S}} (\beta _k ^2 - \lambda h_p^2
P_p )^{ - 1} }, \label{equ18}
\end{equation}
where $\overline{\mathcal{S}}=\{k:\gamma_k=1,\forall k\}$.

It's still very difficult to calculate $X$ and $\gamma_k$ since
$\mathcal{S}$ is determined by $\gamma_k$. In what follows, we
will introduce an iterative algorithm to the problem. It is
observed from (\ref{equ14}) that, when $\lambda$ is small enough,
$\gamma_k$ satisfies $0\le\gamma_k\le 1$ for all $k$. In that
case, $\mathcal{S}=\{1,2,\cdots,K\}$ and $X$ can be easily
calculated by (\ref{equ18}). In addition, given that $\mathcal{S}$
is not changed, both $X$ and $\gamma_k$ increase with the
increasing of $\lambda$. In the case that some $\gamma_k$ turns to
one, $\mathcal{S}$ should be updated, so are $X$ and $\gamma_k$.

\begin{figure} \centering
\includegraphics[width=0.45\textwidth] {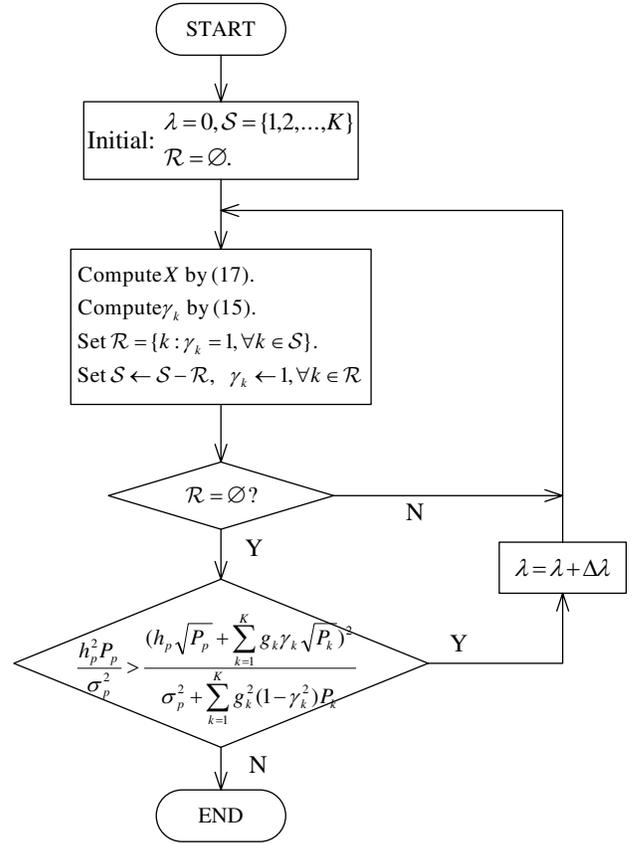}
\caption {Flow chart of the proposed algorithm} \label{fig5}
\end{figure}

The flow chart of the proposed algorithm is depicted in
Fig.\ref{fig5}, where $\Delta\lambda$ is a step size. The
algorithm starts searching from $\lambda=0$ and
$\mathcal{S}=\{1,2,\cdots,K\}$. For each $\lambda$, the algorithm
first computes $X$ by (\ref{equ18}) using the former
$\mathcal{S}$, and computes $\gamma_k(k=1,2,...,K$) by
(\ref{equ16}). If $\mathcal{S}$ is changed, $X$ and $\gamma_k$
should be calculated according to the updated $\mathcal{S}$. The
algorithm converges when $\gamma_k$ satisfies (\ref{equ12}).
%
%
%
%
%
%
%
%
%
%
%
%
%
%

\section{Conclusion}
In this letter, we first study the capacity region of cognitive
multiple access channel under the constraint that the transmission
of primary user is not interfered by cognitive users. Then, we
investigate the optimization of sum-rate capacity on the capacity
region. The optimization is formulated as a nonlinear problem and
an iterative algorithm is introduced.

\end{document}